\documentclass{mn2e}

\usepackage{rotating}
\usepackage{epsfig}

\begin{document}

\def\ltsima{$\; \buildrel < \over \sim \;$}
\def\simlt{\lower.5ex\hbox{\ltsima}}
\def\gtsima{$\; \buildrel > \over \sim \;$}
\def\simgt{\lower.5ex\hbox{\gtsima}}
\def\ls{{_<\atop^{\sim}}}
\def\lax{{_<\atop^{\sim}}}
\def\gs{{_>\atop^{\sim}}}\def\gax{{_>\atop^{\sim}}}
\def\cgs{ ${\rm erg~cm}^{-2}~{\rm s}^{-1}$ } 
 
\title[The HELLAS survey -- VI]
{The $Beppo$SAX High Energy Large Area Survey (HELLAS) - VI. 
The radio properties}

\author [P. Ciliegi et al.] {P. Ciliegi$^1$, C.  Vignali$^2$ , A. 
Comastri$^1$, F. Fiore$^3$, F. La Franca$^4$  and G.C. Perola$^4$ \\ ~ \\
$^1$ INAF - Osservatorio Astronomico di Bologna, Via Ranzani 1, I--40127
Bologna, Italy  \\ 
$^2$ Department of Astronomy and Astrophysics, The Pennsylvania State University, 525 Davey Lab, University Park, PA 16802, USA \\
$^3$ INAF - Osservatorio Astronomico di Roma, Via Frascati 33, I-00040, 
Monteporzio Catone, Italy \\
$^4$ Dipartimento di Fisica, Universit\'a di Roma Tre, Via della Vasca 
Navale 84, I-00146 Roma, Italy \\
}

\maketitle
\begin{abstract}

We present results of a complete radio follow-up 
obtained with the VLA and ATCA radio telescopes down to 
a 6 cm flux limit of about 0.3 mJy (3 $\sigma$) of all the 
147 X-ray sources detected in the $Beppo$SAX HELLAS survey. 
We found 53 X-ray/radio likely associations, corresponding to 
about one third of the X-ray sample. Using the two 
point spectral index  $\alpha_{ro}$=0.35 we divided all the 
HELLAS X-ray sources in radio quiet and radio loud. We have 
26 sources classified as radio-loud objects, corresponding 
to $\sim$18\% of the HELLAS sample. 
In agreement with 
previous results, the identified radio-loud 
sources are associated mainly with Type 1 AGNs  with 
L$_{5-10~keV} \gax 10^{44}$ erg/s,  while all 
the identified Type 2 AGNs and Emission Line Galaxies are 
radio quiet objects with L$_{5-10~keV} \lax 10^{44}$ erg/s. 
The analysis of the radio spectral index 
suggests that Type 1 AGNs have a mean radio spectral index 
($<\alpha_{AGN1}>$=0.25$\pm$0.1) flatter than Type 2 AGNs 
and Emission Line Galaxies ($<\alpha_{AGN2}>$=0.69$\pm$0.11). 
This result is in agreement with the idea that 
the core-dominated radio emission from 
Type 1 AGNs is self-absorbed, while 
in AGN2 and Emission Line Galaxies the radio emission take 
place on larger physical scale, 
without self-absorption. 
 
\end{abstract}
 
 
\section{Introduction}

Radio follow-up of X-ray sources have played an important 
role in the optical identification program of the X-ray 
sources by providing position accurate to $\sim$ 1$^{\prime\prime}$. 
This was particular true for the X-ray sources detected during 
the $Einstein$,  ASCA and $Beppo$SAX 
X-ray missions for which the typical positional 
error is a circle of about 1 arcmin radius. 
A radio detection within the X-ray error box  gives  
fundamental information on the position of the X-ray sources. In fact, 
since the majority of the bright extragalactic X-ray sources 
are associated with AGN, a radio 
source within the X-ray error box is physically associated to the X-ray 
source with a very high probability.  For this reason, many of 
the optical identification programs of the X-ray sources have made an 
intensive use of the radio data (see for example Stocke et al. 1991 for the 
identification of the X-ray sources of the $Einstein$ Extended Medium 
Sensitivity Survey (EMSS) and Akiyama et al. 2000 for the identification 
of the X-ray sources in the ASCA Large Sky Survey (LSS) ).

Moreover, the radio data coupled with the optical and X-ray 
photometry allow us to compute the broad-band two point spectral indices 
$\alpha_{ro}$ and  $\alpha_{ox}$ providing valuable information 
on the nature of the X-ray source population even in the absence of 
optical spectroscopy (Stocke et al. 1991). 

A complete radio follow-up of X-ray selected samples  is also an 
important tool to study the differences between AGNs that 
are strong radio-sources (radio-loud, RL) and those that are radio-quiet 
(RQ). Although the two classes have similar spectral index distributions 
(SEDs) outside the radio band (Elvis et al. 1994), their luminosity 
functions show differences in all the bands in which they have been 
studied (La Franca et al. 1994). 
In the optical band, using the PG sample of optically selected 
AGN, Padovani (1993) has shown that the shapes of the luminosity functions 
for RL and RQ are different. Similar results have been obtained by 
Della Ceca et al. 1994 and Ciliegi et al. 1995 studying the X-ray 
luminosity function (XLF) of RL and RQ separately. In particular 
Della Ceca et al. (1994)  found 
a flattening of the XLF of the RL sample for $L_x \leq10^{44.5}$ erg 
s$^{-1}$. As a result the expected fraction of RL AGNs 
is a function of the X-ray flux limit in X-ray surveys. They predict 
that this fraction is $\sim$ 13 per cent for $f_x(0.3-3.5~keV) 
\sim 2\times10^{-13}$ \cgs 
and decreases to $\sim$2.5 per cent for   $f_x(0.3-3.5~keV)
\sim 2\times10^{-15}$ \cgs.

Radio follow-ups of X-ray selected samples seems to confirm this prediction. 
In fact, while shallow X-ray sample like the $ASCA$ Large Sky Survey 
(Akiyama et al. 2000) with a flux limit of $f_x(2-10~keV) 
\sim 1\times10^{-13}$ \cgs
shows a fraction of RL around 10 per cent, deep 
($f_x(0.5-2.0~keV) \simeq 5-10\times10^{-15}$ \cgs)  
ROSAT samples show a fraction of RL between  2 and  4 per cent
(Ciliegi et al. 1995, de Ruiter et al. 1997, Zamorani 
et al. 1999). 
 
In this paper we report the results of the radio follow-up of all the 147 
X-ray sources detected by the $Beppo$SAX-MECS instrument 
in the framework of the High Energy LLarge Area Survey (HELLAS). 
This survey has observed about 85 deg$^2$ of the sky in the 5-10 keV band 
down to a flux of 4-5 $\times$ 10$^{-14}$ \cgs. The whole survey 
and catalogue is described by Fiore et al. (2001), while the synthesis 
models for the X-ray background and the correlation with the soft X-rays 
have been investigated by Comastri et al. (2001) and Vignali et al. (2001). 
Finally, the spectroscopic identification
 of the HELLAS sources and the study of their 
evolution have been presented by La Franca et al. (2002, hereafter LF02).

\section{Radio Observations} 

The 20 HELLAS sources with a declination further south than $-$40 deg 
have been observed with the Australia Telescope Compact Array 
(ATCA) while the 127 sources with DEC$>-$40 deg have been 
observed with the Very Large Array (VLA). For these latter sources  
a complete covering at 20 cm down to the 5 $\sigma$ flux limit of 
2.5 mJy is already available with the NRAO/VLA Sky Survey (NVSS, Condon 
et al. 1998) while the FIRST survey (Faint Images of the Radio Sky at 
Twenty centimeters, White et al. 1997) is available only for 27 HELLAS 
sources (5 $\sigma$ limit of $\sim$ 1 mJy).  

In order to obtain information also on the radio spectral 
properties of the HELLAS sources 
we adopted the following  strategy. 
All the 147 HELLAS sources have been observed at 6 cm down to 
a 1 $\sigma$ flux limit of $\sim$ 0.10-0.25 mJy. For the 20 HELLAS sources 
observed with the ATCA, we take advantage of the fact that the 6 and 
3 cm receivers of the ATCA share a common feed-horn and we observed 
simultaneously also at 3 cm, obtaining a 3cm flux limit of 
 $\sim$ 0.22 mJy  (1 $\sigma$ level). 

The wavelength of  6 cm and the flux limit reached (0.10-0.25 mJy) 
can be considered a good
compromise between a deep radio survey and the necessity of avoiding 
strong contamination from spurious radio sources within the X-ray 
error box. The expected number of 6 cm sources  is in fact 
N(S)=(0.42$\pm$0.05)(S/30)$^{-1.18\pm0.19}$ where N(S) is the number of
sources per arcmin$^2$ with a flux density $>$S $\mu$Jy (Fomalont et al. 
1991). Considering for the X-ray error box a circle of 1 arcmin radius 
(see below) and a 3 $\sigma$ limit of 0.3 mJy, we expect that only 0.1 
radio source lies just by chance within the HELLAS error box. Going deeper 
in the radio flux  (reaching, for example, a 3 $\sigma$ limit of 0.05 mJy) 
will increase the number of chance coincidence within the HELLAS error 
box to 0.7.  On the other hand, at 20 cm the situation is  more
critical, since at 0.3 mJy the number of  chance coincidence expected 
within an HELLAS error box is $\sim$ 0.3.

\subsection{The ATCA observations} 

The ATCA observations of the 20 HELLAS sources were performed 
on June 1999. They were made with the ATCA simultaneously at two 
different frequencies : 4.848 and 8.640 GHz (referred to as 6 and 3 cm 
in the rest of the paper). The synthesized beam (full width at half power) 
is 2 arcsec at 6 cm and 1 arcsec at 3 cm.  The primary flux density 
calibrator was PKS B1934$-$638, whose flux densities at different 
frequencies are incorporated directly in the calibration software. 
The data were calibrated
and reduced using the ATCA reduction package {\tt MIRIAD} (Multi-channel 
Image Reconstruction Image Analysis and Display). 
For each field  a 512$\times$512 pixel image was constructed, 
with a pixel size of 1 arcsec at 6 cm and of 0.3 arcsec at 3 cm. 
The minimum root 
mean square (rms) noise obtained in each field is $\sim$ 0.25 mJy 
at 6 cm and $\sim$ 0.22 mJy at 3 cm (1 $\sigma$ level).

\subsection{The VLA observations} 
 
The VLA observations were performed on 11 April 2000 
at 4.885 GHz (6 cm) in C configuration. With this configuration 
and frequency, the synthesized beam size is $\sim$ 4 arcsec. 
All the data were analyzed with the NRAO {\tt AIPS} reduction package. 
The data were calibrated using 3C286 as primary flux density 
calibrator. As for the ATCA  data, for each field  a 
512$\times$512 pixel image ($\sim$ 8.5$\times$8.5 arcmin$^2$) 
was constructed with a pixel size of 1 arcsec. 
For the majority of the fields the 1 
$\sigma$ noise obtained in the central area 
is comparable to the expected one ($\sim$ 0.1 mJy). There are however 
some fields 
in which the noise is slightly higher due to problems 
during data acquisitions 
or due to the presence of a nearby strong radio source.

\section{X-ray/radio associations} 

Using our 6 cm maps, we searched for radio sources within the X-ray error box 
of all the 147 HELLAS sources published in Fiore et al. 2001. 
The HELLAS error box has been assumed to be of 90 arcsec, in order to be 
absolutely conservative in the cross-correlation process, even though 
it must be noted that on average the $Beppo$SAX position of the HELLAS 
sources are better defined than 90 arcsec (see Appendix I of Fiore et al. 2001
for a detailed discussion of the position accuracy of the HELLAS sources). 
For the low number of high-Galactic latitude fields with neither a  
target nor a known X-ray source in the same field of view the 
error box has been 
assumed to be of 2 arcmin since no correction to the astrometry was
possible (Vignali 2001).   

For all the  radio sources detected within the HELLAS error box we have 
computed the probability of random association with the X-ray source. 
Assuming that the radio sources belong to a Poissonian distributed 
population of sources, 

$P_{XR}=1-e^{-N(S)\pi d^2}$ \\
gives the probability to have a random association within a distance $d$ 
(distance between the X-ray position and the radio position of the possible 
counterpart) with a radio source having a flux greater than S. N(S) is the 
expected number of source with flux greater than that of the possible 
counterpart (see Section 2). 

As first step, P$_{XR}<$0.01 was 
chosen as a convenient threshold to accept a 
radio/X-ray association. We have 32 
radio/X-ray associations with  P$_{XR}<$0.01.  However, we  note that 
this threshold represents only a starting point in the radio X-ray 
association process since further and more tightening constraints on 
the most likely associations come from the  position of the optical 
counterparts of the HELLAS sources reported in LF02. 
Using a subsample of 118 HELLAS sources (enclosed in a region 
with DEC$<$79 deg and outside 5$^h<$RA$<6.5^h$ and  17$^h<$RA$<20^h$) 
they report the optical identification of 61 HELLAS : 37 Type 1 AGN, 
9 Type 2 AGN, 5 narrow emission-line galaxies (ELG), 6 Cluster, 2 BL Lac, 1 
Radio Galaxy and 1 Star. Of these 61 sources, 24 have a radio counterpart 
within 5 arcsec and with P$_{RO}<$ 0.0002
(P$_{RO}$ has been calculated as  P$_{XR}$ using the distance between 
the radio and optical position as value for $d$).
Assuming that all the 24 radio/optical associations are real associations
(as suggested by their low  P$_{RO}$ values), 
we analyzed their  P$_{XR}$ values. For 19 radio/optical associations
we found P$_{XR}<$0.01, 4 have 
0.01$<$P$_{XR}<$0.1 and one has P$_{XR}=$0.217. 
The large value of P$_{XR}$ for the 5 sources with  P$_{XR}>$0.01
is due to the fact that for these  sources 
the distance between the radio/optical position and the X-ray 
position is greater than 60 arcsec (see LF02 for a 
discussion on the procedure of the optical identification of the HELLAS 
sources).  In the light of these results, we increase our threshold 
for the radio X-ray associations to P$_{XR}=$0.1 ($i.e.$ we expect that about 
10 per cent of the proposed identifications may be spurious positional 
coincidences). 
With this new threshold we found 53 X-ray/radio associations. 
We do not have  X-ray sources with two (or more) radio counterparts both with 
P$_{XR}<$0.1.  

\begin{sidewaystable*}



\label{tab1}

\tiny

\begin{tabular}{lrrcccrrrrcrcrrrr}
\\\hline\hline

 $BeppoSAX$ Position    &   Offx &  Fx   &  R   & ID & z    &  F8.6        &       F5         & F1.4           &$\alpha_r$ &Radio Position             &$\Delta_{XR}$   &   P$_{XR}$ & $\Delta_{RO}$   &   P$_{RO}$ & $\alpha_{ro}$ & $\alpha_{ox}$ \\ 
 RA(2000) DEC(2000)  &($\prime$) &       &      &    &      &  (mJy)       &    (mJy)         & (mJy)          &         &RA(2000) DEC(2000)         & ($\prime\prime$)& $\times10^{-2}$ & ($\prime\prime$)& $\times10^{-6}$ \\
\hline
00 26 36.5 $-$19 44  13 &   13.2 &  3.39 & 18.1  & 2 & 0.238 &              & $<$0.30          &  $<$2.5        &         &                            &       &      &     &      &$<$0.09 &  1.23 \\
00 27 09.9 $-$19 26  19 &    6.2 &  1.83 & 17.7  & 1 & 0.227 &              & $<$0.30          &  $<$2.5        &         &                            &       &      &     &      &$<$0.06 &  1.40 \\
00 27 43.9 $-$19 30  29 &   11.0 &  1.58 &       &   &       &              & $<$0.30          &  $<$2.5        &         &                            &       &      &     &      &        &       \\
00 45 49.6 $-$25 15  13 &   23.7 &  3.26 & 17.5  & 2 & 0.111 &              & $<$0.30          &  $<$2.5        &         &                            &       &      &     &      &$<$0.05 &  1.33 \\
00 48 05.8 $-$25 04  32 &   14.9 &  1.48 &       &   &       &              & $<$0.40          &  $<$2.5        &         &                            &       &      &     &      &        &       \\
01 18 03.9 +89 20  12   &    8.3 &  0.87 &       &   &       &              & $<$0.30          &  $<$2.5        &         &                            &       &      &     &      &        &       \\
01 21 56.8 $-$58 44  05 &   15.6 &  2.57 & 16.8  & 2 & 0.118 &$<$0.65       & $<$0.75          &                &         &                            &       &      &     &      &$<$0.07 &  1.48 \\
01 34 14.3 $-$29 45  41 &   14.2 &  1.50 &       &   &       &              & $<$0.35          &  $<$2.5        &         &                            &       &      &     &      &        &       \\
01 34 28.6 $-$30 06  34 &   12.2 &  1.33 &$>$20  &   &       &              &  0.41$\pm$0.05   &  4.3$\pm$0.5   &  1.89   & 01 34 25.61 $-$30 05 50.90 & 57.9  & 5.46 &     &      &$>$0.26 &$<$1.10\\
01 34 33.3 $-$29 58  38 &    6.0 &  0.87 & 18.0  & 1 & 2.217 &              & $<$0.35          &  $<$2.5        &         &                            &       &      &     &      &$<$0.10 &  1.48 \\
01 34 49.6 $-$30 02  34 &    5.4 &  0.71 &       &   &       &              & $<$0.35          &  $<$2.5        &         &                            &       &      &     &      &        &       \\
01 35 30.2 $-$29 51  21 &    8.4 &  0.90 & 17.7  & 1 & 1.344 &              & $<$0.35          &  $<$2.5        &         &                            &       &      &     &      & $<$0.07&  1.52 \\
01 40 08.9 $-$67 48  13 &    8.0 &  2.83 & 12.4  & 0 & 0.000 &$<$0.65       & $<$0.75          &                &         &                            &       &      &     &      &$<-$0.26&  2.14 \\
01 53 03.9 +89 12  20   &    2.4 &  0.54 &       &   &       &              & $<$0.35          &  $<$2.5        &         &                            &       &      &     &      &        &       \\
02 42 01.8 +00 00  46   &   10.0 &  1.47 & 18.6  & 1 & 1.112 &              & $<$0.45          &  $<$2.5        &         &                            &       &      &     &      &$<$0.16 &  1.30 \\
02 42 09.4 +00 02  29   &    7.5 &  0.68 &18.4$^u$   &   &       &              & 1.65$\pm$0.20    &  $<$2.5        &$<$0.33  & 02 42 14.53 +00 02 51.95   & 80.3  & 2.07 & 1.1 & 3.71 &  0.25  &  1.46 \\
03 08 08.5 +89 08  41   &    4.9 &  0.65 &       &   &       &              & $<$0.30          &  $<$2.5        &         &                            &       &      &     &      &        &       \\
03 08 19.0 +02 46  27   &   18.5 &  5.09 &$>$20  &   &       &              &                  &  7.4$\pm$0.5   &         & 03 08 21.80 +02 46 34.50   & 42.6  & 0.29 &     &      &$>$0.33 &$<$0.88\\
03 15 45.0 $-$55 29  26 &   14.5 &  2.65 & 17.9  & 1 & 0.464 &$<$0.65       & $<$0.75          &                &         &                            &       &      &     &      &$<$0.15 &  1.31 \\
03 17 32.4 $-$55 20  12 &   21.0 &  4.10 & 17.5  & 1 & 0.406 &$<$0.65       & $<$0.75          &                &         &                            &       &      &     &      &$<$0.12 &  1.30 \\
03 33 09.6 $-$36 19  40 &   15.2 &  3.98 & 17.5  & 3 & 0.308 &              &                  & 13.6$\pm$0.6   &         & 03 33 12.54 $-$36 19 47.51 & 36.3  & 0.10 & 4.1 & 9.90 &  0.28  &  1.30 \\
03 34 07.4 $-$36 04  22 &    5.1 &  1.95 & 20.1  & 1 & 0.904 &              &                  &  $<$2.5        &         &                            &       &      &     &      &$<$0.30 &  1.02 \\
03 36 51.3 $-$36 15  57 &   14.7 &  3.72 & 17.7  & 1 & 1.537 &              & 583$\pm$42       & 501$\pm$15   & -0.12   & 03 36 54.01 $-$36 16 05.11 & 33.7  & 0.00 & 0.9 & 0.00 &  0.67  &  1.28 \\
04 32 27.9 +05 13  05   &   15.0 &  2.06 &       &   &       &              & $<$0.75          & $<$2.5         &         &                            &       &      &     &      &        &       \\
04 37 14.5 $-$47 30  58 &   16.0 &  2.68 & 17.3  & 1 & 0.142 &$<$0.65       & $<$0.75          &                &         &                            &       &      &     &      &$<$0.11 &  1.40 \\
04 38 47.9 $-$47 29  06 &   20.1 &  4.69 & 20.5  & 1 & 1.453 &94.99$\pm$1.5 & 132.3$\pm$2.2    &                &  0.58   & 04 38 47.01 $-$47 28 00.91 & 65.7  & 0.01 & 1.1 & 0.02 &  0.73  &  0.81 \\
05 02 15.5 +12 04  07   &   20.9 &  7.41 &       &   &       &              & $<$0.30          &  $<$2.5        &         &                            &       &      &     &      &        &       \\
05 15 13.7 +01 08  07   &    7.6 &  1.26 &$>$20  &   &       &              & 0.48$\pm$0.05    &  $<$2.5        &$<$1.33  & 05 15 11.83 +01 08 20.98   & 31.3  & 1.35 &     &      &$>$0.27 &$<$1.11\\
05 20 48.3 $-$45 42  00 &    9.6 &  5.90 &$>$20  &   &       &$<$0.65       & 4.90$\pm$0.10    &                &         & 05 20 44.78 $-$45 41 27.33 & 49.3  & 0.22 &     &      &$>$0.46 &$<$0.85\\
05 48 41.3 $-$60 52  18 &   19.5 &  2.42 &       &   &       &$<$0.65       & $<$0.75          &                &         &                            &       &      &     &      &        &       \\
05 50 00.2 $-$61 02  22 &    5.9 &  0.90 &14.5$^u$   &   &       &$<$0.65       & 0.97$\pm$0.10    &                &$>$0.70  & 05 50 06.61 $-$61 01 23.26 & 75.0  & 3.35 & 2.6 &40.99 &$-$0.08 & 2.01  \\
05 52 06.1 $-$60 59  48 &   11.1 &  1.23 &       &   &       &$<$0.65       & $<$0.75          &                &         &                            &       &      &     &      &        &       \\
05 52 51.3 $-$60 57  18 &   17.1 &  2.06 &       &   &       &$<$0.65       & $<$0.75          &                &         &                            &       &      &     &      &        &       \\
06 23 56.6 $-$69 21  13 &    6.1 &  1.33 &17.4$^u$&  &       &$<$0.65       & 0.59$\pm$0.10    &                &$>-$0.17 & 06 23 50.91 $-$69 20 08.07 & 71.5  & 5.42 & 0.5 &2.72  &  0.09  &  1.50 \\
06 25 31.3 $-$69 19  09 &    6.4 &  1.42 &       &   &       &$<$0.65       & $<$0.75          &                &         &                            &       &      &     &      &        &       \\
06 46 39.3 $-$44 15  35 &   16.6 &  4.27 & 16.6  & 1 & 0.153 &              & $<$0.50          &  $<$2.5        &         &                            &       &      &     &      &$<$0.02 &  1.43 \\
06 46 42.7 $-$44 32  29 &   19.0 &  2.68 &       &   &       &              & $<$0.50          &  $<$2.5        &         &                            &       &      &     &      &        &       \\
06 55 39.6 +79 10  48   &    4.6 &  0.75 &       &   &       &              & $<$0.30          &  $<$2.5        &         &                            &       &      &     &      &        &       \\
07 21 29.6 +71 14  04   &    8.1 &  0.84 & 17.7  & 1 & 0.232 &              & $<$0.40          &  $<$2.5        &         &                            &       &      &     &      &$<$0.08 &  1.53 \\
07 41 40.3 +74 14  57   &   22.6 & 30.66 & ...   & 4 & 0.216 &              & 1.50$\pm$0.15     & 22.7$\pm$1.6   &  2.19   & 07 41 44.70 +74 14 39.76   & 24.9  & 0.22 & 2.0 &14.50 &        &       \\
07 41 45.2 +74 26  23   &   13.1 &  3.74 &       &   &       &             & $<$0.35          &  $<$2.5        &        &                            &       &      &      &      &          &         \\
07 43 09.1 +74 29  19   &    7.2 &  6.05 & 16.4  & 1 & 0.312 &             & $<$0.35          &  $<$2.5        &        &                            &       &      &      &      &$<-$0.02  &   1.40  \\
08 37 37.2 +25 47  48   &   12.1 &  2.63 & 16.9  & 1 & 0.077 &             & 1.07$\pm$0.10    &  1.18$\pm$0.15 &  0.09  & 08 37 37.03 +25 47 50.47   & 3.4   & 0.00 & 1.8  &17.50 &    0.10  &   1.46  \\
08 38 59.9 +26 08  13   &   23.0 & 16.38 & 15.3  & 5 & 0.048 &             & 1.89$\pm$0.15    &  5.19$\pm$0.15 &  0.81  & 08 38 59.27 +26 08 13.07   & 8.5   & 0.02 & 1.3  & 4.66 &    0.03  &   1.40  \\
09 46 05.3 $-$14 02  59 &   15.8 &  2.82 &       &   &       &             & $<$0.35          &  $<$2.5        &        &                            &       &      &      &      &          &         \\
09 46 17.9 $-$14 10  27 &   10.9 &  2.04 &18.7$^a$&  &       &             & 0.37$\pm$0.05    &  $<$2.5        &$<$1.54 & 09 46 21.89 $-$14 10 50.29 & 62.5  & 7.12 & 3.0  &170.16&    0.15  &   1.23  \\
09 46 32.8 $-$14 06  15 &   16.4 &  3.22 &       &   &       &             & $<$0.30          &  $<$2.5        &        &                            &       &      &      &      &          &         \\
10 29 19.1 +50 48  15   &   17.5 &  5.78 &       &   &       &             & $<$0.35          &  $<$1.0        &        &                            &       &      &      &      &          &         \\
10 32 15.8 +50 51  03   &   10.4 &  3.13 & 15.9  & 1 & 0.174 &             & $<$0.30          &  $<$1.0        &        &                            &       &      &      &      &$<-$0.07  &   1.59  \\
10 34 43.1 +39 29  18   &    8.4 &  1.11 &       &   &       &             & $<$0.30          &  $<$1.0        &        &                            &       &      &      &      &          &         \\
10 34 52.0 +39 40  12   &    5.3 &  1.41 & 16.5$^u$&   &       &             & $<$0.30          &  1.51$\pm$0.15 &$>$1.30 & 10 34 56.41 +39 39 40.78   & 59.8  & 3.61 & 0.5  & 1.46 & $-$0.03  &   1.63  \\
10 52 45.4 +57 30  42   &    4.0 &  0.83 & 20.9$^l$& 6$^l$& 0.708$^l$&          & 16.09$\pm$0.15   & 65.0$\pm$2.30  &  1.12  & 10 52 37.39 +57 31 04.09   & 68.2  & 0.10 & 0.0  & 0.00 &    0.62  &   1.04  \\
10 54 19.8 +57 25  09   &   13.4 &  2.62 & 18.5  & 2 & 0.205 &             & 0.50$\pm$0.05    &  1.44$\pm$0.15 &  0.83  & 10 54 21.14 +57 25 44.63   & 37.2  & 1.81 & 0.8  & 8.48 &    0.16  &   1.22  \\
10 54 21.7 +57 36  24   &   16.6 &  2.13 & 17.6$^u$&   &       &             & 37.30$\pm$0.15   & 191.4$\pm$6.70 &  1.32  & 10 54 26.24 +57 36 48.19   & 43.8  & 0.02 & 1.2  & 0.12 &    0.44  &   1.39  \\
11 01 46.4 +72 26  11   &   22.3 &  7.29 & 16.7  & 1 & 1.460 &             & 864.1$\pm$1.30   & 1245.0$\pm$37.4&  0.29  & 11 01 48.92 +72 25 38.00   & 34.9  & 0.00 & 0.5  & 0.00 &    0.63  &   1.32  \\
11 02 37.2 +72 46  38   &   20.7 &  7.87 & 15.1  & 1 & 0.089 &             & $<$0.30          & $<$2.5         &        &                            &       &      &      &      &$<-$0.13  &   1.56  \\
11 06 14.0 +72 43  16   &    8.5 &  1.67 & 18.5  & 1 & 0.680 &             & $<$0.30          & $<$2.5         &        &                            &       &      &      &      &$<$0.12   &   1.29  \\
11 07 04.9 $-$18 16 28  &    7.7 &  1.23 &       &   &       &             & $<$0.30          & $<$2.5         &        &                            &       &      &      &      &          &         \\
11 18 11.9 +40 28  33   &    4.2 &  0.85 & 18.7  & 1 & 0.387 &             & $<$0.35          & $<$1.0         &        &                            &       &      &      &      &$<$0.15   &   1.37  \\
11 18 46.2 +40 27  39   &    4.8 &  1.39 & 18.5  & 1 & 1.129 &             & $<$0.30          & $<$1.0         &        &                            &       &      &      &      &$<$0.12   &   1.32  \\
11 34 52.7 +70 23  09   &   15.5 &  3.77 &       &   &       &             & $<$0.30          & $<$2.5         &        &                            &       &      &      &      &          &         \\
11 56 39.2 +65 17  57   &    5.7 &  0.75 &       &   &       &             & $<$0.30          & $<$2.5         &        &                            &       &      &      &      &          &         \\
11 57 01.7 +65 27  24   &   15.4 &  2.14 &       &   &       &             & $<$0.30          & $<$2.5         &        &                            &       &      &      &      &          &         \\
12 04 07.6 +28 08  30   &   16.1 &  5.10 & ...   & 4 & 0.167 &             & $<$0.30          & $<$1.0         &        &                            &       &      &      &      &          &         \\
12 17 45.1 +47 29  55   &   16.4 &  3.24 & $>$20 &   &       &             & 0.41$\pm$0.07    & $<$1.0         &$<$0.72 & 12 17 51.07 +47 30 13.58   & 63.3  & 6.49 &      &      &$>$0.26   & $<$0.95 \\
12 17 50.3 +30 07  08   &   19.8 &  3.54 & 14.0  & 3 & 0.237 &             & 384.8$\pm$3.1    & 571.6$\pm$21.4 &  0.32  & 12 17 52.08 +30 07 00.33   & 24.3  & 0.00 & 1.7  & 0.02 &    0.36  &   1.86  \\
12 18 55.0 +29 58  12   &   12.7 &  1.98 & 18.6  & 2 & 0.176 &             & $<$0.30          & $<$1.0         &        &                            &       &      &      &      &$<$0.13   &   1.25  \\
12 19 21.6 +47 11  07   &    9.4 &  1.32 & $>$20 &   &       &             & $<$0.30          & 9.08$\pm$0.15  &$>$2.75 & 12 19 21.27 +47 12 13.46   & 66.5  & 0.69 &      &      &$>$0.23   & $<$1.10 \\
12 19 45.7 +47 20  42   &    7.5 &  1.17 & 19.3  & 1 & 0.654 &             & 1.21$\pm$0.05    & 2.90$\pm$0.50  &  0.70  & 12 19 52.36  +47 20 58.65  & 69.7  & 2.24 & 0.9  & 3.78 &    0.29  &   1.23  \\
12 22 06.8 +75 26  17   &    6.5 &  2.46 & ...   & 4 & 0.240 &             & 1.60$\pm$0.05    & $<$2.5         &$<$0.36 & 12 22 06.50 +75 26 14.9    &  2.2  & 0.00 & 0.1  & 0.00 &          &         \\
12 29 23.7 +01 51  38   &   14.1 &  1.61 &       &   &       &             & $<$4.0           & $<$24.5        &        &                            &       &      &      &      &          &         \\
12 40 26.0 $-$05 13  20 &   11.7 &  3.13 & 18.8  & 1 & 0.300 &             & $<$0.30          & $<$1.0         &        &                            &       &      &      &      &$<$0.14   &   1.14  \\
12 40 29.6 $-$05 07  46 &   16.8 &  1.92 & 15.2  & 5 & 0.008 &             & $<$0.30          &  1.30$\pm$0.15 &$>$1.18 & 12 40 36.89 $-$05 07 52.80 & 109.0 & 21.7 & 4.3  &108.16&    0.14  &   1.77  \\
12 54 28.0 +59 21  01   &   24.3 &  6.37 &       &   &       &             & $<$0.30          & $<$2.5         &        &                            &       &      &      &      &          &         \\
12 55 16.6 $-$05 39  22 &   16.3 &  2.20 &       &   &       &             & $<$3.00          & $<$5.6         &        &                            &       &      &      &      &          &         \\
12 56 09.9 $-$05 54  30 &    7.6 &  0.91 &       &   &       &             & $<$3.00          & $<$5.6         &        &                            &       &      &      &      &          &         \\

\hline
\end{tabular}
\end{sidewaystable*}

\begin{sidewaystable*}

\setcounter{table}{0}

\footnotesize

\tiny

  \label{tab1_con}
\begin{tabular}{lrrcccrrrrcrcrrrr}
\\\hline\hline

 $BeppoSAX$ Position    &   Offx &  Fx   &  R   & ID & z    &  F8.6        &       F5         & F1.4           &$\alpha_r$ &Radio Position             &$\Delta_{XR}$   &   P$_{XR}$ & $\Delta_{RO}$   &   P$_{RO}$ & $\alpha_{ro}$ & $\alpha_{ox}$ \\ 
 RA(2000) DEC(2000)  &($\prime$) &       &      &    &      &  (mJy)       &    (mJy)         & (mJy)          &         &RA(2000) DEC(2000)         & ($\prime\prime$)& $\times10^{-2}$ & ($\prime\prime$)& $\times10^{-6}$ \\
\hline
13 04 24.3 $-$10 23  53 &    4.4 &  1.28 &       &   &       &             & $<$1.00          & $<$2.5         &        &                            &       &      &      &      &          &         \\
13 04 38.2 $-$10 15  47 &    5.9 &  1.43 & 20.1  & 1 & 2.386 &             & $<$0.30          & $<$2.5         &        &                            &       &      &      &      &$<$0.24   &   1.07  \\
13 04 45.1 $-$05 33  37 &    7.5 &  1.26 & $>$20 &   &       &             & 0.58$\pm$0.05    & $<$1.5         &$<$0.76 & 13 04 44.22 $-$05 33 40.21 & 13.5  & 0.20 &      &      &$>$0.29   &$<$1.11  \\
13 05 32.3 $-$10 32  35 &   22.0 & 19.27 & 14.9  & 1 & 0.278 &             & 1149.6$\pm$1.0  & 711.3$\pm$21.3 & $-$0.39  & 13 05 33.02 $-$10 33 19.12 & 45.4  & 0.00 & 0.9  & 0.00 &    0.52  &   1.44  \\
13 05 36.5 $-$05 43  30 &   22.7 &  6.42 &$>$20  &   &       &             & 2.10$\pm$0.11    & 6.30$\pm$0.50  &  0.88 & 13 05 36.10 $-$05 42 00.31 & 89.7  & 1.94 &     &      &$>$0.39  & $<$0.84  \\
13 36 34.3 $-$33 57  47 &   21.8 &  3.18 & 10.5  & 6 & 0.013 & 440$^n$     & 1886$^n\pm$132   & 2730.4$^n\pm$80&  0.30 & 13 36 39.01 $-$33 57 58.00 & 60.0  & 0.00 & 0.1 &  0.00&   0.23  &   2.41  \\
13 38 34.1 +48 21  05   &    4.3 &  1.24 &       &   &       &             & $<$0.30          & $<$1.0         &       &                            &       &      &     &      &         &         \\
13 42 47.9 +00 21  09   &   16.9 &  2.48 & 19.8$^a$&   &       &             & 29.5$\pm$0.5     & 96.8$\pm$3.5   &  0.96 & 13 42 46.03  +00 20 28.30  & 49.4  & 0.03 & 2.3 &  0.57&   0.59  &   1.03  \\
13 42 59.3 +00 01  38   &   20.6 &  3.25 & 18.7  & 1 & 0.804 &             &  $<$0.30         & $<$1.0         &       &                            &       &      &     &      &$<$0.14  &   1.15  \\
13 48 20.8 $-$30 11  06 &   14.4 &  2.18 & 15.3 5& 2 & 0.128 &             & 4.68$\pm$0.13    & 10.0$\pm$0.6   &  0.61 & 13 48 19.50 $-$30 11 54.01 & 50.9  & 0.24 & 2.0 &  3.79&   0.10  &   1.74  \\
13 48 24.3 $-$30 25  47 &   14.8 &  3.15 &       &   &       &             & $<$0.50          & $<$2.5         &       &                            &       &      &     &      &         &         \\
13 48 37.9 $-$30 09  11 &   12.3 &  1.59 &       &   &       &             & $<$0.50          & $<$2.5         &       &                            &       &      &     &      &         &         \\
13 48 45.4 $-$30 29  36 &   14.4 &  5.11 & 17.1  & 1 & 0.330 &             & $<$0.30          & $<$2.5         &       &                            &       &      &     &      &$<$0.02  &   1.32  \\
13 50 09.4 $-$30 19  55 &   10.8 &  5.08 & 16.5 4& 5 & 0.074 &             & 0.69$\pm$0.11    & $<$2.5         &$<$1.04& 13 50 15.37 $-$30 20 09.58 & 78.6  & 5.44 & 0.6 &  3.26&   0.04  &   1.41  \\
13 53 54.6 +18 20  33   &   17.5 &  6.82 & 17.3  & 1 & 0.217 &             & 0.43$\pm$0.05    & $<$1.0         &$<$0.68& 13 53 54.43  +18 20 16.59  & 16.6  & 0.43 & 0.7 &  7.76&   0.06  &   1.24  \\
13 55 54.1 +18 13  35   &   19.8 &  6.14 &       &   &       &             & $<$0.45          & $<$1.0         &       &                            &       &      &     &      &         &         \\
14 11 58.7 $-$03 07  02 &   20.7 &  3.93 &       &   &       &             & $<$0.35          & $<$1.0         &       &                            &       &      &     &      &         &         \\
14 17 12.5 +24 59  28   &   12.9 &  0.69 & 19.5  & 1 & 1.057 &             & $<$0.30          & $<$1.0         &       &                            &       &      &     &      &$<$0.19  &   1.29  \\
14 18 31.1 +25 11  07   &    8.6 &  6.11 & ...   & 4 & 0.240 &             &  12.8$\pm$0.20   & 38.0$\pm$1.5   &  0.88 & 14 18 32.38   +25 12 00.99 & 56.7  & 0.09 & 4.0 &  4.62&         &         \\
14 38 30.1 +64 30  25   &   11.2 &  2.57 & 17.2$^u$& &       &             & 0.65$\pm$0.05    & $<$2.5         &  1.08 & 14 38 26.71   +64 28 59.76 & 88.0  & 7.25 & 0.5 &  0.00&   0.09  &   1.42  \\
14 48 21.8 $-$69 20  30 &   11.7 &  4.25 & 17.0$^u$& &       &     $<$0.65 & 0.95$\pm$0.10    &                &$>$0.66& 14 48 25.59 $-$69 19 52.07 & 42.9  & 1.14 & 2.1 & 27.41&   0.10  &   1.37  \\
15 19 39.9 +65 35  46   &   13.9 &  9.43 & 14.4  & 2 & 0.044 &             & 1.08$\pm$0.20    & $<$2.5         &  0.68 & 15 19 33.69   +65 36 00.80 & 41.2  & 0.90 & 2.8 & 41.88& $-$0.08 &   1.63  \\
15 28 46.0 +19 45  10   &    4.7 &  1.65 & 20.3$^u$& &       &             & 1.16$\pm$0.05    & 1.26$\pm$0.16  &  0.09 & 15 28 44.59   +19 44 34.94 & 40.3  & 0.79 & 0.4 & 0.79 &   0.36  &   1.02  \\
15 28 47.3 +19 39  10   &    5.0 &  1.57 & 20.3  & 1 & 0.657 &             &   $<$0.30        & $<$1.0         &       &                            &       &      &     &      &$<$0.25  &   1.03  \\
16 26 56.8 +55 13  24   &   14.2 &  3.15 &       &   &       &             & $<$0.30          & $<$1.0         &       &                            &       &      &     &      &         &         \\
16 26 59.9 +55 28  20   &   10.5 & 12.09 & ...   & 4 & 0.130 &             &  $<$ 0.30        & $<$1.0         &       &                            &       &      &     &      &         &         \\
16 34 10.7 +59 37  44   &    6.9 &  0.50 &       &   &       &             &  $<$ 0.30        & $<$2.5         &       &                            &       &      &     &      &         &         \\
16 34 11.0 +59 48  15   &    5.2 &  0.48 & 17.6$^u$& &       &             & 0.87$\pm$0.15    & $<$2.5         &$<$0.85&  16 34 12.53 +59 49 13.75  & 59.9  & 2.44 & 1.5 & 15.51&   0.14  &   1.64  \\
16 34 11.8 +59 45  29   &    3.4 &  0.84 & 19.0  & 2 & 0.341 &             &  $<$0.30         & $<$2.5         &       &                            &       &      &     &      &$<$0.16  &   1.33  \\
16 49 57.9 +04 53  32   &   19.8 &  9.45 & ...   & 4 & 0.154 &             &  $<$0.30         & $<$2.5         &       &                            &       &      &     &      &         &         \\
16 50 40.1 +04 37  17   &   24.5 & 12.25 & 14.6  & 2 & 0.031 &             &  1.63$\pm$0.10   & $<$2.5         &$<$0.34&  16 50 42.77  +04 36 18.39 & 70.9  & 1.63 & 1.1 &  3.98& $-$0.03 &   1.56  \\
16 52 12.5 +02 11  29   &   15.8 &  2.36 &       &   &       &             &  $<$0.30         & $<$2.5         &       &                            &       &      &     &      &         &         \\
16 52 38.0 +02 22  18   &    4.6 &  0.67 & 20.7  & 1 & 0.395 &             & $<$0.30          & $<$2.5         &       &                            &       &      &     &      &$<$0.28  &   1.11  \\
16 54 41.1 +40 02  10   &   17.6 &  6.32 & 14.9$^u$& &       &             & 36.3$\pm$0.35    & 80.7$\pm$2.9  &  0.64 & 16 54 43.03  +40 02 46.20  & 42.4  & 0.02 & 2.3 & 0.45 &   0.24  &   1.62  \\
17 40 10.7 +67 42  50   &   17.7 &  3.28 & 18.3$^a$& &       &             & 5.57$\pm$0.06    &  4.10$\pm$0.40 & $-$0.25 & 17 40 22.32  +67 41 38.14  & 97.6  & 0.73 & 0.7 & 0.37 &   0.34  &   1.21  \\
17 42 36.3 +68 00  44   &   10.6 &  2.76 &$>$20  &   &       &             & 2.71$\pm$0.09    & 7.3$\pm$0.50   &  0.80 & 17 42 50.51  +67 59 33.50  & 106.5 & 2.02 &     &      &$>$0.41  &$<$0.98  \\
17 50 25.4 +60 56  01   &   11.7 &  2.10 &$>$20  &   &       &             & 2.11$\pm$0.11    & 4.6$\pm$0.40   &  0.63 & 17 50 38.81 +60 56 52.50   & 110.4 & 2.91 &     &      &$>$0.39  &$<$1.02  \\
17 51 30.3 +61 00  43   &    6.7 &  0.55 &       &   &       &             & $<$0.30          & $<$ 2.5        &       &                            &       &      &     &      &         &         \\
17 52 38.3 +61 05  47   &    3.2 &  0.36 &       &   &       &             & $<$0.30          & $<$ 2.5        &       &                            &       &      &     &      &         &         \\
17 53 49.0 +60 59  52   &   12.7 &  1.10 &$>$20  &   &       &             & 0.47$\pm$0.10    & $<$2.5         &$<$1.35& 17 53 47.71 +60 58 58.76   & 54.0  & 4.07 &     &      &$>$0.27  &$<$1.13  \\
18 03 51.8 +61 10  21   &    2.9 &  0.60 &$>$20  &   &       &             & 39.0$\pm$0.50    & 132.0$\pm$5.0  &  0.98 &  18 03 47.38 +61 09 22.20  & 66.9  & 0.03 &     &      &$>$0.62  &$<$1.23  \\
18 15 17.5 +49 44  51   &    9.8 &  3.48 &       &   &       &             & $<$0.30          & $<$2.5         &       &                            &       &      &     &      &         &         \\
18 18 58.6 +61 14  42   &   18.6 &  2.07 &$>$20  &   &       &             & 2.54$\pm$0.06    & $<$2.5         &$<-$0.01& 18 18 51.06 +61 14 18.43  & 59.3  & 0.68 &     &      &$>$0.40  &$<$1.03  \\
18 19 18.4 +60 56  06   &    3.5 &  1.52 &       &   &       &             & $<$0.30          & $<$2.5         &       &                            &       &      &     &      &         &         \\
18 19 35.5 +60 58  46   &    2.1 &  0.49 &       &   &       &             & $<$0.30          & $<$2.5         &          &                            &       &      &     &      &         &        \\
18 19 39.5 +60 53  26   &    3.7 &  0.83 &       &   &       &             & $<$0.30          & $<$2.5         &          &                            &       &      &     &      &         &        \\
18 36 11.3 $-$65 07  20 &   22.8 &  4.44 &       &   &       & $<$1.0      & $<$1.0           &                &          &                            &       &      &     &      &         &        \\
20 42 47.6 $-$10 38  30 &   21.1 &  5.32 & 17.9  & 1 & 0.363 &             & $<$0.30          & $<$2.5         &          &                            &       &      &     &      &$<$0.08  &   1.19 \\
20 44 34.8 $-$10 27  34 &   15.3 &  1.99 & 17.7  & 1 & 2.755 &             & 4.30$\pm$0.17    & 8.8$\pm$0.5    &   0.58   & 20 44 34.95 $-$10 28 09.09 & 35.2  & 0.13 & 2.5 & 6.54 &   0.28  &   1.39 \\
21 22 59.2 +89 01  58   &   15.3 &  4.77 &$>$20  &   &       &             & 19.50$\pm$0.17    & 60.6$\pm$1.9   &   0.91   & 21 21 36.16 +89 02 16.36   & 27.9  & 0.01 &     &      &$>$0.57  &$<$0.89 \\
21 38 08.9 $-$14 33  13 &    7.0 &  2.34 &       &   &       &             & $<$0.30          & $<$2.5         &          &                            &       &      &     &      &         &        \\
21 42 46.7 +89 35  30   &   23.5 & 10.47 &       &   &       &             & $<$0.30          & $<$2.5         &          &                            &       &      &     &      &         &        \\
21 59 52.1 +88 54  53   &   17.8 &  6.85 &       &   &       &             & $<$0.30          & $<$2.5         &          &                            &       &      &     &      &         &        \\
22 03 00.5 $-$32 04  18 &   16.5 &  2.83 &       &   &       &             & $<$0.35          & $<$2.5         &          &                            &       &      &     &      &         &        \\
22 26 30.3 +21 11  56   &   13.7 &  3.91 & 17.6  & 1 & 0.260 &             & $<$0.40          & $<$2.5         &          &                            &       &      &     &      &$<$0.08  &   1.29 \\
22 31 49.6 +11 32  08   &   15.2 &  1.94 &       &   &       &             & $<$1.00          & $<$2.5         &          &                            &       &      &     &      &         &        \\
22 41 22.1 +29 42  41   &   15.6 &  2.98 &$>$20  &   &       &             & 0.84$\pm$0.11    & $<$2.5         & $<$0.88  &  22 41 27.56 +29 43 33.88  & 88.6  & 5.49 &     &      &$>$0.31  &$<$0.96 \\
22 42 51.5 +29 35  32   &    7.6 &  1.02 &       &   &       &             & $<$1.00          & $<$2.5         &          &                            &       &      &     &      &         &        \\
22 44 11.4 +29 51  14   &   23.1 &  3.95 &       &   &       &             & $<$1.00          & $<$2.5         &          &                            &       &      &     &      &         &        \\
23 02 30.1 +08 37  06   &   17.6 &  2.67 &       &   &       &             & $<$0.30          & $<$2.5         &          &                            &       &      &     &      &         &        \\
23 02 36.2 +08 56  42   &   10.1 &  3.17 &       &   &       &             & $<$0.30          & $<$2.5         &          &                            &       &      &     &      &         &        \\
23 06 59.2 +08 48  40   &    4.1 &  1.02 &       &   &       &             & $<$0.30          & $<$2.5         &          &                            &       &      &     &      &         &        \\
23 15 36.4 $-$59 03  40 &    8.3 &  2.03 & 11.2 2& 5 & 0.044 &11.01$\pm$0.10& 15.40$\pm$0.10   &                &   0.58   & 23 15 46.72 $-$59 03 15.79 & 83.4  & 0.16 & 1.9 & 0.84 & $-$0.10 &   2.38 \\
23 16 09.8 $-$59 11  24 &    6.1 &  1.32 &       &   &       & $<$0.65     & $<$0.75          &                &          &                            &       &      &     &      &         &        \\
23 19 22.1 $-$42 41  50 &   21.4 &  5.67 & 16.5  & 5 & 0.101 &  $<$0.65    & $<$0.75          &                &          &                            &       &      &     &      &$<$0.05  &   1.39 \\
23 27 28.7 +08 49  30   &    7.1 &  0.55 & 18.5  & 1 & 0.154 &             & $<$0.35          & $<$2.5         &          &                            &       &      &     &      &$<$0.13  &   1.48 \\
23 27 37.1 +08 38  56   &    7.9 &  1.48 &       &   &       &             & $<$0.35          & $<$2.5         &          &                            &       &      &     &      &         &        \\
23 29 02.4 +08 34  39   &   20.0 &  2.87 & 20.3 2& 1 & 0.953 &             & 197.8$\pm$0.40   & 172.9$\pm$5.2  &  $-$0.11   & 23 29 05.79 +08 34 16.45   & 55.0  & 0.00 & 0.5 & 0.00 &   0.78  &   0.92 \\
23 31 55.6 +19 38  34   &   17.0 &  3.75 & 18.8  & 1 & 0.475 &             & $<$0.35          & $<$2.5         &          &                            &       &      &     &      &$<$0.16  &   1.11 \\
23 55 32.7 +28 35  11   &    7.7 &  1.72 &       &   &       &             & $<$0.75          & $<$2.5         &          &                            &       &      &     &      &         &        \\
23 55 53.3 +28 36  05   &   12.0 &  4.17 & 17.9  & 1 & 0.731 &             & 286.7$\pm$0.50   & 681.0$\pm$20.4 &   0.70   &  23 55 54.1  +28 35 58.00  & 13.6  & 0.00 & 2.6 & 0.05 &   0.63  &   1.23 \\
\hline							 
							 
\end{tabular}

\vspace{1mm}

$^a$ magnitude from the APM catalogue \\
$^l$ Lehmann et al (2000) \\ 
$^n$ data from the NASA/IAPC Extragalactic Database (NED) \\
$^u$ magnitude from the USNO catalogue \\

\normalsize						 
\end{sidewaystable*}					 

\setcounter{table}{1}
							 		
\subsection{Optical counterparts of the X-ray/radio associations}

Starting from the radio position of the  53 X-ray/radio associations, 
we searched for optical counterparts within 5 arcsec from the 
radio position using the optical positions of the 61 HELLAS sources 
identified by LF02, the USNO-A2.0 
\footnote {http://archive.eso.org/skycat/servers/usnoa}
optical catalogue, 
the APM 
\footnote{http://www.ast.cam.ac.uk/$\sim$apmcat/} optical catalogue  and 
the Nasa Extragalactic Database (NED). 
As said above, 24  X-ray/radio associations have been identified with 
sources in LF02 (10 Type 1 AGN, 4 Type 2 AGN, 2 BL LAC, 3 Clusters, 4 ELGs 
and 1 Radio galaxy), 1 has been identified with a z=0.708 Radio galaxy in 
the Lockman Hole using NED (see table 2  source 116 in 
Lehmann et al. 2000 for a description of this source), 13 have an optical 
(R band) identification in the USNO and/or APM catalogue while  15 
X-ray/radio associations do not have  an optical identification brighter than 
R=20.  The list of all the 147 HELLAS sources with their most probable 
optical and radio counterparts is shown in Table 1. The first three columns 
show the X-ray position, the offset in arcmin in the $Beppo$SAX field 
and the 5-10 keV X-ray flux in units of 
10$^{-13}$ \cgs (from Fiore et al. 2001). Columns IV, V and VI 
give the R band magnitude (from LF02 if not otherwise specified), 
spectroscopic identification (1 for Type I AGN, 2 for Type II AGN, 3 for 
BL LAC, 4 for Cluster, 5 for ELG and 6 for Radio galaxies) and 
redshift from LF02.  Columns VII, VIII, and IX give the radio flux 
 at 3cm, 6 cm and 20 cm, respectively. All the 
20 cm data are from the NVSS catalogue version 2.17 
if not otherwise specified. For the X-ray sources without radio counterparts, 
we report at 3 and 6 cm the 3 $\sigma$ upper limit of our maps while at 
20 cm we report the limit of the radio catalogue used (NVSS or FIRST). 
Column X gives the radio spectral index  
calculated using the available frequencies
(mainly 6 and 20 cm). Column XI gives the position of the radio sources 
obtained from our 6 cm map (except when the source is detected only at 20 cm). 
The typical positional error of 6 cm sources is below 1 arcsec. 
Columns XII and XIII 
give the distance between the radio and the X-ray position 
and the probability $P_{XR}$ of a random association.  
Columns XIV and XV give the distance between the radio and the optical 
position and the probability $P_{RO}$ of a random association.  Finally 
columns XVI and XVII give the two point spectral indices 
$\alpha_{ro}$ and  $\alpha_{ox}$ defined as 
$\alpha_{ro} = log(F_{5GHz}/F_{2500})/5.38 $ and 
$\alpha_{ox} = log(F_{2500}/F_{2~keV})/2.605 $ where $F_{5GHz}$ is 
the radio flux at 6 cm obtained from the present observation or by 
extrapolating the 20 cm flux density with a power-law slope with 
$\alpha_r$=0.7, $F_{2500}$ is the optical flux at 2500\AA ~derived 
from the 
R magnitude with an optical slope of $\alpha_o$=0.5 while $F_{2~keV}$ is 
the monochromatic X-ray flux at 2 keV derived from the measured $Beppo$SAX 
5-10 keV flux and assuming an energy spectral index $\alpha_x$=0.6 
(Fiore et al. 2001). 

\subsubsection{Comparison with the optical identification reported in LF02} 

As said above, 24 of the 61 sources with an optical identification reported 
in LF02 have a radio counterpart within 5 arcsec. In addition to this 
information, it is interesting to note that no other radio sources have 
been found within their X-ray error box, as well as no radio sources have 
been detected in the X-ray error box of the remaining 37 sources.  
Moreover, among the 13 X-ray sources  classified as empty fields 
down to R=21 in LF02, only two (0134$-$3006 and 1304$-$0533) have a radio 
counterpart (see Table 1). These two radio sources have not an optical
identification brighter than R=21 within 5 arcsec from their radio 
position, in agreement with their classification as empty field. 

These results, under the hypothesis that a radio 
source within the X-ray error box is physically associated to the X-ray 
source with a very high probability, strengthen the goodness of the 
optical identification reported in LF02.

\subsection{Radio-Loud and Radio-Quiet Classification} 

\begin{figure}
\centerline{
\epsfig{file=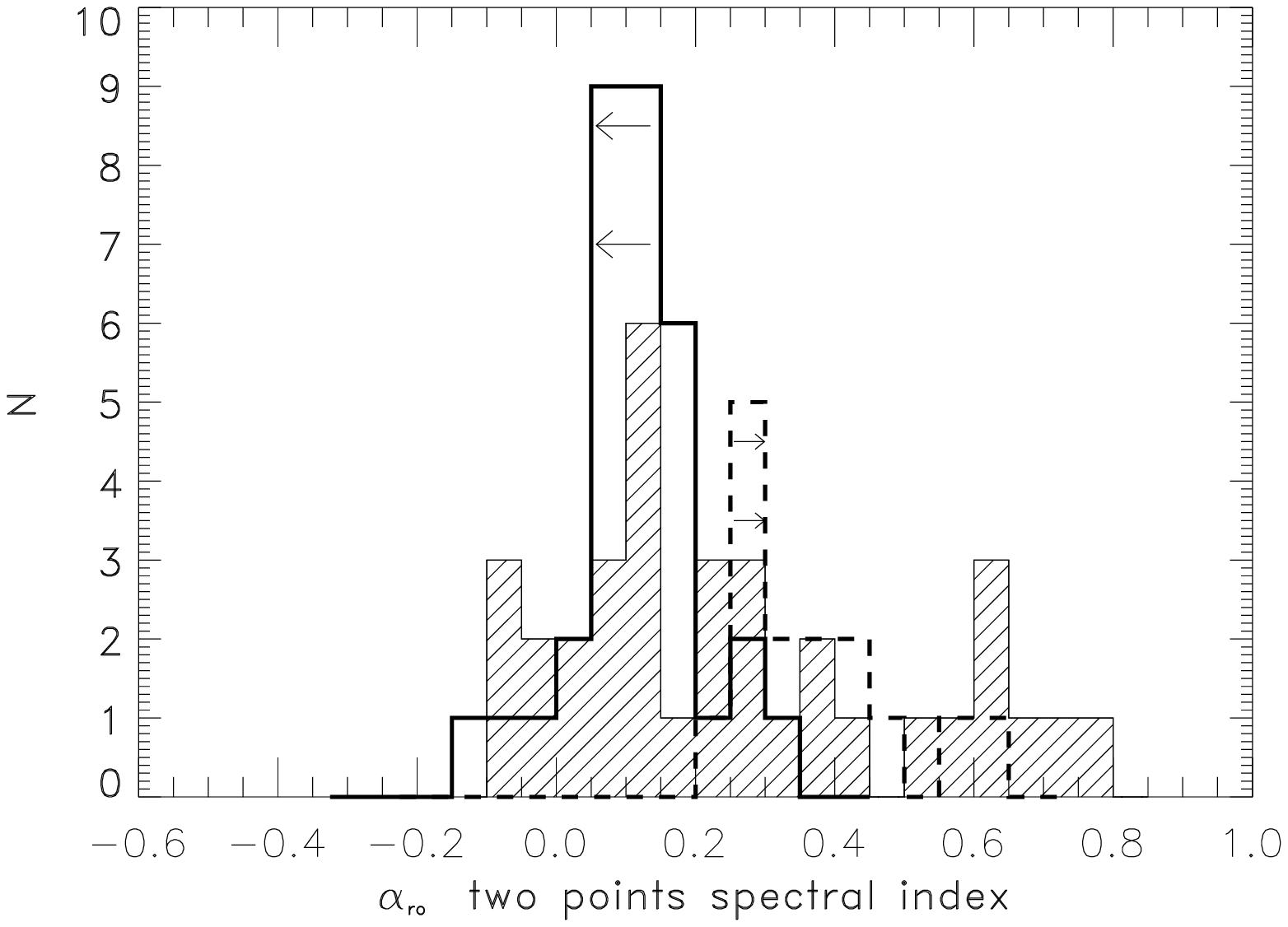, width=9cm}
}
\caption{ Ditribution of the two point spectral index $\alpha_{RO}$ 
for the HELLAS sources (sources detected both in radio and optical band: {\em shaded histogram};  
3$\sigma$ upper limits (X-ray sources with an optical counterpart 
but without a radio counterpart): {\em open solid line histogram}; 
3$\sigma$ lower limits (X-ray sources with a radio counterpart 
but without an optical counterpart):  {\em open dashed line histogram}). 
}
\label{Aro_distribution}
\end{figure}

The two point spectral index $\alpha_{ro}$ has been commonly used 
to discriminate between radio-loud and radio-quiet objects. 
We were able to calculate the 
$\alpha_{ro}$ value  for 
a total of 84 sources: 55 HELLAS sources identified in LF02 
(we did not calculated 
the two point spectral indicies for the 6 clusters due to the indetermination 
of their optical magnitude), for the z=0.708 Radio galaxy in 
the Lockman Hole (see above), for the 13 X-ray/radio associations that 
have been identified with an USNO and/or an APM source and for the 15 
X-ray/radio associations without an optical counterpart (in the latter 
case using an optical upper limit of R=20). The $\alpha_{ro}$ distribution 
of these 84 sources is shown in Figure~\ref{Aro_distribution} 
(sources detected both in radio and optical band: {\em shaded histogram};  
3$\sigma$ upper limits (X-ray sources with an optical counterpart 
but without a radio counterpart): {\em open solid line histogram}; 
3$\sigma$ lower limits (X-ray sources with a radio counterpart 
but without an optical counterpart):  {\em open dashed line histogram}). 

\begin{figure}
\centerline{
\epsfig{file=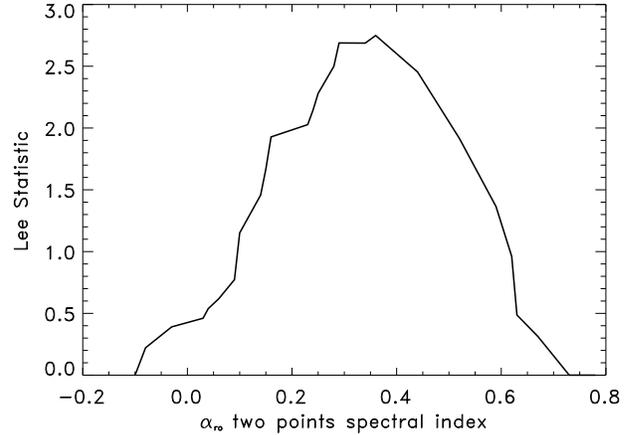, width=9cm}
}
\caption{Lee statistic for the distribution of the two point spectral index
 $\alpha_{ro}$. The position of the maximum is the value of the variable 
which best divides the two distributions, $i.e.$, the radio-quiet 
and radio-loud objects.  
}
\label{lee}
\end{figure}

The values of $\alpha_{ro}$ used in literature to discriminate 
between radio-loud and radio-quiet objects are 0.20 (see Elvis et al. 1994, 
Giommi, Menna \& Padovani 1999) or 0.35 (Zamorani et al. 1981, 
Della Ceca et al. 1994). 
To choose between these two values, we applied to our  $\alpha_{ro}$
distribution (Figure~\ref{Aro_distribution}) the Lee statistic 
(see Fitchett \& Merrit 1988 and Lee 1979 for details). The Lee statistic 
is a test for a  bimodality of a distribution. The location of 
the maximum corresponds to the value of the variable which best separates 
the two subsamples. The results of the Lee statistic are shown in 
Figure~\ref{lee}. We find that the Lee parameter reaches it maximum for 
$\alpha_{ro} \simeq $0.35 - 0.36. This results has been obtained using 
only the distribution of the sources detected both in radio and 
optical band (shaded histogram in Figure~\ref{Aro_distribution}). However 
similar results are obtained treating the upper and lower limits 
as detections.  Consequently the value of   
$\alpha_{ro}$=0.35 has been used to separate the RL  from 
the RQ objects. 

\begin{figure}
\centerline{
\epsfig{file=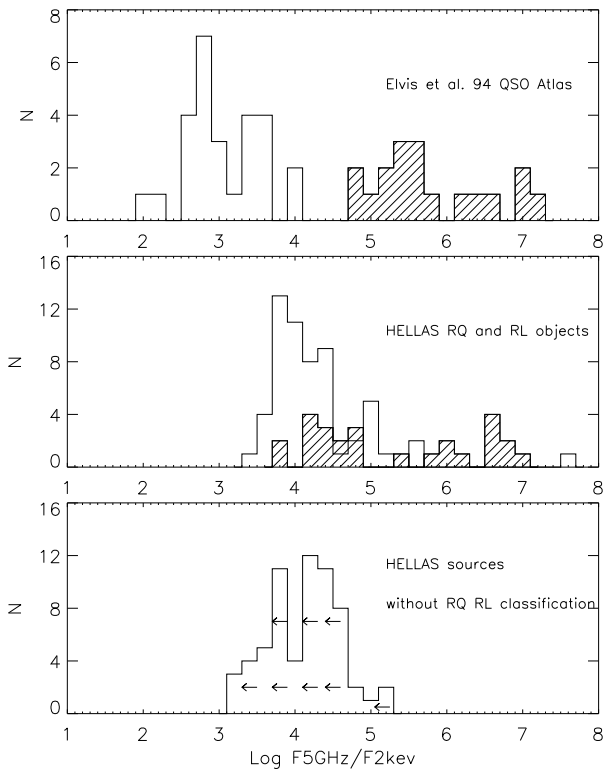, width=9cm}
}
\caption{ {\em Top panel}: radio-to-X-ray ratio distribution for the 29
Radio-Quiet (empty histogram) and 18 Radio-Loud (shaded histogram) objects in 
the atlas of quasars from Elvis et al. (1994).  
{\em Middle panel}: As above for the 58 Radio-Quiet and 26 Radio-Loud objects 
in the HELLAS survey. 
{\em Bottom Panel}: Distribution of the radio-to-X-ray ratio upper limit
for the 63 HELLAS sources for which only an X-ray detection and a radio 
upper limit are available. 
}
\label{x_radio_ratio}
\end{figure}

We have 18 RL sources with 
$\alpha_{ro}\geq0.35$ (11 have been detected both in the  radio and 
optical bands and 7 have a lower limit on $\alpha_{ro}$ greater than 0.35), 
58 RQ sources (24 detected both in the radio and optical bands, 34 
with an   $\alpha_{ro}$ upper limit lower than 0.35) while for 8 sources 
it was impossible to give a reliable classification due to the fact that 
they have an   $\alpha_{ro}$ lower limit lower than 0.35. 
However, given the fact that 
their  $\alpha_{ro}$ lower limit are close to the threshold of 0.35 
(the lowest value is  $\alpha_{ro}>$0.23, see Table 1 and the open dashed line 
histogram in Figure~\ref{Aro_distribution}) we assumed these 
8 sources to be RL objects. Over 84 sources analyzed we have therefore
26 RL and 58 RQ objects. For the remaining 63 HELLAS sources it was impossible 
to calculate the two point spectral index  $\alpha_{ro}$ 
(or at least a lower or an upper 
limit) due to the lack of a radio and optical identification, $i.e.$ for 
these sources only an X-ray detection and a radio upper limit is available.
However also the upper limit on their radio-to-X-ray ratio can be used 
to obtain useful informations on the radio nature of these source.  
In Figure ~\ref{x_radio_ratio} 
(top panel)  we report the distribution of the 
radio-to-X-ray ratio for the RQ and  RL objects in the atlas of 
quasars published by Elvis et al. (1994) while in the the other two panels we 
report the distribution of the 84 HELLAS sources for which it was possible to 
obtain a classification as RQ  and RL  object (middle panel) 
and the distribution of the radio-to-X-ray ratio upper limits for the  63 
HELLAS sources without a classification in RL and RQ (bottom panel).  
Although the atlas of 47 quasars from Elvis et al. (1994) is an 
heterogeneous sample of objects selected mainly from the PG 
(Schmidt and Green 1983), 3C and Parkes catalogues, it gives a useful 
reference for the  radio-to-X-ray ratio values for RQ and RL objects. 
As shown in figure ~\ref{x_radio_ratio} the RQ and RL objects show 
a significantly different distribution of their radio-to-X-ray ratio. 
A Kolmogorov-Smirnov (KS) test of the  radio-to-X-ray ratio distributions for 
the HELLAS RQ and RL (reported in the middle panel of 
figure ~\ref{x_radio_ratio}) shows that the two distributions are 
drawn from the same population with a probability 
P$_{KS}<5\times10^{-4}.$ The same result has been obtained performing
 a KS test 
between the distribution of the RL and that of the 63 sources for which
 only and X-ray detection and a radio upper limit is available. Viceversa,
a KS test between this latter and the distribution of the 
HELLAS RQ objects indicate that the hypothesis that the two distribution 
are drawn from the same population can not be excluded (P$_{KS}$=0.69). 
It is therefore reasonable to assume that all the 63 sources 
for which only and X-ray detection and a radio upper limit is available 
are radio-quiet objects. Under this hypothesis, the fraction of radio-loud 
objects in the HELLAS sample is $\sim$18 per cent (26 RL over a total of 
147 sources).

\section{Discussion} 

\subsection{Radio, Optical and X-ray properties} 

\begin{figure}
\centerline{
\epsfig{file=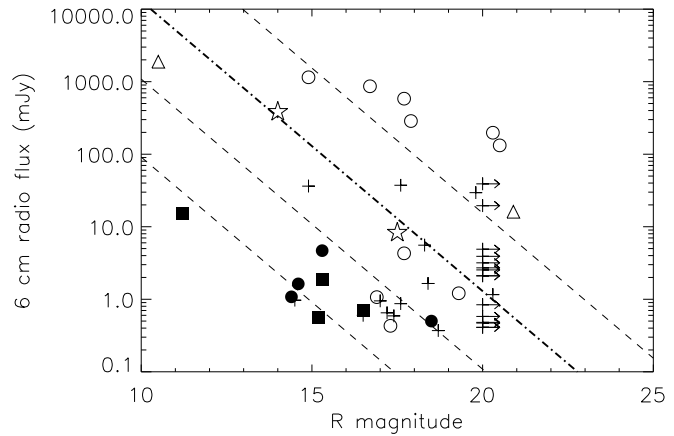, width=9cm}
}
\caption{ The R-band magnitude versus the 6 cm radio flux. Open circles : 
Type 1 AGN; filled circles : Type 2 AGN; Open Stars : BL LAC objects; 
 Filled Squares : Emission Line Galaxies; Open Triangles : Radio 
Galaxies; Crosses : Unidentified HELLAS sources. The lines represent different 
two point spectral index $\alpha_{ro}$, corresponding to 
 $\alpha_{ro}$ = $-$0.05, 0.15, 0.35 and 0.55 (from left to right). 
Objects above the 
dot dashed line ($\alpha_{ro}$=0.35) are RL sources. 
}
\label{Rmag_f6cm}
\end{figure}

\begin{figure}
\centerline{
\epsfig{file=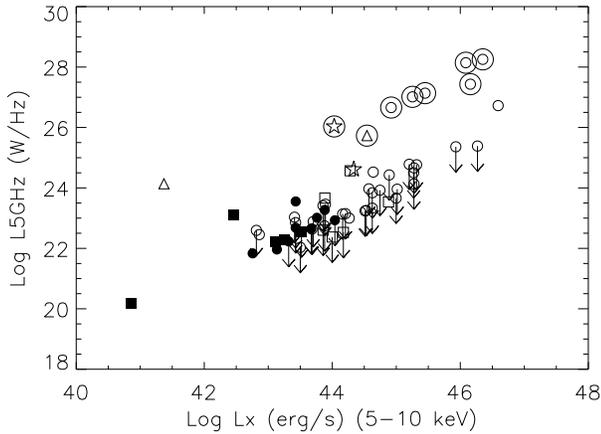, width=9cm}
}
\caption{ Radio versus hard X-ray 
luminosity for the HELLAS sample. The X-ray luminosity
has been corrected for the photoelectric absorption. 
The sources with the encircled 
symbol are the objects classified as radio-loud sources on the basis 
of their  radio to optical ratio $\alpha_{ro}>$0.35.  
Symbols as in Figure~\ref{Rmag_f6cm}. The six 
empty squares  show the sources identified as cluster. 
 }
\label{Lr_Lx}
\end{figure}

To examine the general properties of the radio/X-ray associations, 
in Figure~\ref{Rmag_f6cm} we show the 6 cm radio flux versus the R-band 
magnitude for the 53 radio/X-ray associations of the HELLAS sample. 
Superimposed are the lines corresponding to constant values 
for the two point spectral index $\alpha_{ro}$. The dot dashed line  
shows the value of $\alpha_{ro}$=0.35 used to separate RL and RQ objects. 
In   Figure~\ref{Lr_Lx}  we plot the 
radio versus the hard (5-10 keV) X-ray luminosities. 
The latter has been  corrected for absorption 
using the hardeness ratio reported in Fiore et al. (2001).

It is clear from Figures~\ref{Rmag_f6cm} and ~\ref{Lr_Lx}  
that  the  
radio-loud objects are associated mainly with Type 1 AGNs  with 
L$_{5-10~keV} \gax 10^{44}$ erg/s,  
while all the  Type 2 AGNs and 
Emission Line Galaxies are radio-quiet sources with  
$\alpha_{ro}$ $\lax$ 0.15  
and 
L$_{5-10~keV} \lax 10^{44}$ erg/s.  

The lack of low luminosity radio-loud sources is well explained by the 
dependence of the radio-loud fraction to the X-ray and optical 
luminosity. The fraction of the radio-loud AGN 
changes, in fact, from $\sim$18\% at L$_x \simeq 4 \times10^{44}$ erg/s
to $\sim$2\% at L$_x \simeq 2 \times10^{43}$ 
in X-ray selected samples (Della Ceca et al. 1994) and from
$\sim$20\%-50\% at M$_B \leq-$24.5 to $\sim$7\%-8\%  at fainter absolute 
magnitude in optically selected samples 
(Padovani 1993, La Franca et al. 1994, Goldschmidt et al. 1999). 

The lack of high X-ray luminosity Type 2 AGN is also a well know effect. 
Only few examples have been reported prior the $Chandra$ mission 
(see, for example, Almaini et al. 1995, Boyle et al. 1998, 
Franceschini et al. 2000).  
However, even in the deeper $Chandra$ and XMM surveys only a couples of 
sources considered as the prototype of Type 2 QSO 
(L$_{2-10} > 10^{44}$, $N_H>10^{23}$ cm$^{-2}$, optical emission 
lines with FWHM $\lax$ 2000 km/s) have been detected 
(Norman et al. 2002, Stern et al. 2002).  
Their space density will be better constrained by the 
forthcoming  deep X-ray ($Chandra$ and XMM) and far-infrared 
surveys (SIRTF). 

\begin{figure}
\centerline{
\epsfig{file=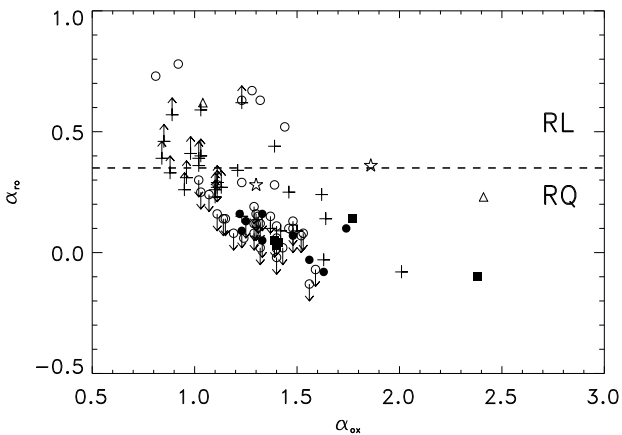, width=9cm}
}
\caption{ Overall energy distributions for the HELLAS sources.
The axes are the radio-to-optical $\alpha_{ro}$ and the X-ray-to-optical
$\alpha_{ox}$ two point spectral indices. The objects plotted are 
 the 84 HELLAS sources for which 
we were able to calculate  $\alpha_{ox}$ and $\alpha_{ro}$. The horizontal 
dashed line shows the value of $\alpha_{ro}$=0.35 used to divided RL and RQ
objects. Symbols as in Figure~\ref{Rmag_f6cm}. 
}
\label{Aro_Aox}
\end{figure}

Finally, in Figure~\ref{Aro_Aox} we show the 
$\alpha_{ox}$-$\alpha_{ro}$ plot for the 84 HELLAS sources for which 
we were able to calculate  $\alpha_{ox}$ and $\alpha_{ro}$. 
From this figure it is evident that the RQ Type 1 AGNs and the Type 2 AGNs 
occupy the same region in the plane $\alpha_{ox}$-$\alpha_{ro}$ and 
then the two classes cannot be distinguished only on the basis of their 
multiband photometric data. On the other hand, the region with  
$\alpha_{ox}$ $\lax$ 1.5 and $\alpha_{ro} \gax$ 0.35 is occupied only by 
Type 1 AGN. We would therefore conclude that all the unidentified sources 
with  $\alpha_{ro}$ $\gax$ 0.30 - 0.35 and $\alpha_{ox} \lax$ 1.5 
are likely to be Type 1 AGN. 

\subsection{Comparison with other X-ray samples}

\begin{table*}
\caption{\bf Summary of the radio/X-ray associations from different X-ray surveys }

\label{tab_rx_ass}

\begin{tabular}{lcccr}
\hline

X-RAY  SURVEY & f$_x$ limit &  f$_r$ limit & fr/f$_{2keV}$ &    
 \% Radio/X    \\
        & erg/sec/cm$^2$ &  mJy  &  ($\times10^3$) &
 associations  \\ \hline
$^1$ Marano ROSAT + 20cm ATCA & 4$\times10^{-15}$ (0.5-2.0 keV) & 0.20 & 335 & (4/50) $\sim$ ~8\%      \\
$^2$ LOCKMAN ROSAT + 6cm VLA  & 1$\times10^{-15}$ (0.5-2.0 keV) & 0.05 & 279 & (8/54) $\sim$ 15\%    \\ 
$^3$ CRSS ROSAT + 20cm VLA    & 2$\times10^{-14}$ (0.5-2.0 keV) & 0.70 & 234 & (7/80) $\sim$ ~9\%     \\
$^4$ EMSS $Einstein$ + 6cm VLA & 8$\times10^{-14}$ (0.3-3.5 keV) & 0.80 & 119  & (167/625) $\sim$ 27\%  \\
$^5$ LSS ASCA + 20cm FIRST      & 1$\times10^{-13}$ (2.0-10~ keV) & 1.00 & 78  & (12/34) $\sim$ 35\%   \\
$^6$ HELLAS $Beppo$SAX + 6cm VLA-ATCA  & 5$\times10^{-14}$ (5.0-10~ keV) & 0.30 & 20  & (53/147) $\sim$ 36\% \\
\hline
\end{tabular}
\footnotesize

     $^1$ Zamorani et al. 1999; $^2$Ciliegi et al. 2003; $^3$ Ciliegi et al. 1995; 
$^4$Stocke et al. 1991;  $^5$Akiyama et al. 2000; $^6$ this work 
\end{table*}


\normalsize

\subsubsection{ The $Einstein ~ Observatory$, ROSAT and ASCA surveys} 

Previous radio follow-up of X-ray selected samples 
have shown different percentages
of radio/X-ray associations. Using 20 cm observations of ROSAT X-ray selected 
sample, Ciliegi et al. (1995), de Ruiter et al. (1997) and Zamorani et al. 
(1999) found only $\sim$ 10\% of radio/X-ray coincidences, while 
Stocke et al. (1991) found a significantly higher fraction (about 27\%) 
for the X-ray sources in the $Einstein$ Extended Medium Sensitivity Survey
(EMSS). For the HELLAS sources we have 53 radio/X-ray associations over 
a total of 147 sources. From the statistical analysis of the 
radio/X-ray associations (see Section 3), we estimate that about 10\% 
($i.e.$ $\sim$ 5 sources) of the proposed  associations 
may be spurious and therefore the percentage with the ``correct'' 
radio/X-ray associations is in the range $\sim$33-36\%. This value is 
comparable to that obtained by Akiyama et al. (2000) in the course of the 
$ASCA$ Large Sky Survey ($\sim$ 35\%), one of the few large area samples  
 selected in the hard X-ray band and with a complete follow-up in the radio 
and optical bands. 

\begin{figure}
\centerline{
\epsfig{file=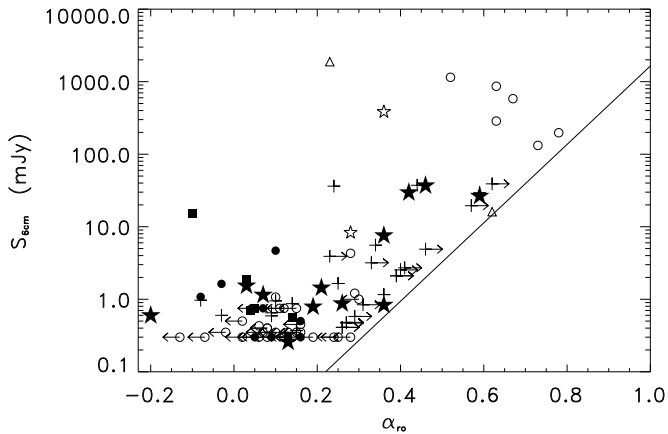, width=9cm}
}
\caption{ The 6 cm radio flux  as function of the two point 
spectral index $\alpha_{ro}$ for the  HELLAS sources  
(symbols as in Figure~\ref{Rmag_f6cm}) and for the 12 radio/X-ray 
associations found by Akiyama et al. (2000) in the ASCA LSS 
survey (filled stars). The solid line shows the limit beyond 
which a source can not be 
detected in the optical band due to the magnitude limit of the 
optical data (R=21 mag). 
}
\label{Radioflux_Aro}
\end{figure}

In Table~\ref{tab_rx_ass} we report a summary of different 
X-ray surveys for which a complete radio follow-up is available. 
All these surveys are ``AGN dominated'' in the sense that the 
large majority of their identification ($\geq$75\%) are associated 
with AGNs. 
For each survey we report the X-ray and radio limits, the radio-to-X-ray 
limit ratio  fr/f$_{2keV}$ 
assuming an X-ray spectral index $\alpha_x$=1.0, 
the percentage of radio/X associations (number of radio/X-ray
association over the total number of X-ray sources in the survey) and the 
fraction of Radio-Loud objects in the radio/X-ray associations. 
As clearly shown in the table, the 
percentage of radio/X-ray associations is a function, as expected,
 of  fr/f$_{2keV}$: the lower this ratio is 
($i.e$ deeper radio data in comparison 
to the X-ray flux limit) the higher is the fraction of X-ray objects with 
a radio counterpart. The top panel of Figure~\ref{x_radio_ratio}
suggests (as proposed also by Akiyama et al. 2000) that this change 
in the percentage of  radio/X-ray 
associations is due to the fact that when  fr/f$_{2keV}$ 
becomes very low we start seeing also  radio-quiet AGNs. 
To check if this is the case, in 
Figure~\ref{Radioflux_Aro} we plot the 6 cm radio flux as function of the 
two point spectral index $\alpha_{ro}$  for the HELLAS sources and 
for the 12 radio/X-ray associations found in the ASCA LSS (using a radio 
spectral index    $\alpha_{r}$=0.7 to convert the radio fluxes from 20 cm 
to 6 cm). The solid line shows the limit beyond which a source can not be 
detected in the optical band due to the magnitude limit of the 
optical data. The line in Figure~\ref{Radioflux_Aro} has been drawn using 
R=21. As clearly shown in figure, going deeper in the radio data we start 
to see the radio-quiet population ($\alpha_{ro}<$0.35). This 
population  is almost 
completely absent at radio flux level greater than 10-20 mJy but becomes 
the dominant population at radio fluxes lower than $\sim$1 mJy. From 
Figure~\ref{Radioflux_Aro} it is also evident that  deep optical 
data are needed in order to detect in the optical band 
radio-loud sources  with radio fluxes lower than a few mJy.  

Moreover it is interesting to note that the 
 3 X-ray surveys with the highest 
percentage of radio/X-ray associations (EMSS, LSS and HELLAS) are all selected 
in the medium or hard X-ray band (see Table~\ref{tab_rx_ass}), while the 3
X-ray surveys with the lowest percentage of  radio/X-ray associations 
(Lockman, Marano and CRSS) are all selected in the soft X-ray band. In 
addition to the lowest radio-to-X-ray flux ratio available in the 
EMSS, LSS and HELLAS surveys, the higher percentage of radio/X-ray 
associations could also be due to the selection in the hard X-ray band. 
In fact, since the radio emission is transparent against obscuring material, 
obscured AGNs are detected in an unbiased way in the radio wavelength 
as well as in the hard X-ray band, but they could be  missed 
in samples selected in the soft X-ray band where obscuration effects 
are much more relevant. To test this hypothesis we calculate the fraction 
of radio/X-ray associations in the HELLAS and LSS surveys using a radio 
flux limits of 3.7 mJy and 3.2 mJy respectively, in order to have an 
 fr/f$_{2keV}$  similar to that of the soft X-ray 
selected samples (see Table~\ref{tab_rx_ass}).  We found a percentage 
of radio/X-ray associations around 20\%, i.e. still higher than that 
found in soft X-ray selected samples. In conclusion the higher fraction 
of radio/X-ray associations found in the EMSS, LSS and HELLAS surveys 
in comparison to the soft X-ray selected samples is probably due 
to a combination of two effects:  deeper radio 
data in comparison to the X-ray flux limit and the selection in harder 
X-ray bands.

\subsubsection{The deep $Chandra$ surveys}

Recently new deep X-ray samples have been obtained thanks to the revolutionary
capabilities  of the $Chandra$ X-ray 
satellite. The deep surveys (Brandt et al. 
2001, Rosati et al. 2002) have reached flux limits at least two orders of 
magnitude lower than the fainter samples listed in Table~\ref{tab_rx_ass}. 
Extensive programs of optical identifications showed that most of the 
sources in these samples are not AGN.
The bulk of the optical identifications in these deep X-ray 
surveys is associated with normal galaxies, the majority of which show 
strong emission lines indicative of star formation activity
(Barger et al. 2001, Brandt et al. 2001,Bauer et al. 2002).  
However, the most surprising finding is the discovery of a sizeable number 
of relatively bright X-ray sources spectroscopically identified with 
early-type ``normal'' galaxies without any signature of nuclear activity 
in the optical spectra (Fiore et al. 2000, Barger et al. 2001). 
These optically normal X-ray luminous galaxies were thought to be relatively 
rare, unusual objects, but during the recent deep X-ray surveys 
obtained with $Chandra$ and XMM we saw a remarkable increase of their number
(Comastri et al. 2002).

The radio properties of these new deep X-ray samples have been 
studied using very deep VLA radio survey (down to few micro Jy  at 
1.4 GHz) obtained in the SSA13 field (Barger et al. 2001) and in the 
HDF-N (Bauer et al. 2002). These studies showed a different behavior 
of these new samples in comparison to the ``AGN-dominated'' samples. 
In fact, in the new deep X-ray surveys a very high percentage 
of radio/X-ray associations (up to 80\% in the SSA13 field, see Barger 
et al. 2001) has been found, 
with the highest matching fraction found among the emission line 
galaxies, which are comprised of apparently normal and starburst galaxies 
at redshift of z$\sim$0.1-1.3 thought to be undergoing recent star formation. 
The nature of the radio/X-ray associations is therefore completely  
different between that found  in the ``AGN dominated'' samples 
reported in Table 2 and that found in the new deep X-ray surveys. 
While the former are associated with an AGN activity, the latter are 
mainly associated with a star forming activity.

\subsection{The radio spectral index} 

Using all the radio data available at 3, 6 and 20 cm we were able 
to calculate the radio spectral index (or an upper/lower limit 
when the source is detected in only one radio band) for a total 
of 50 HELLAS sources.  Since the majority of the spectral indices have 
been calculated using the 6 cm fluxes obtained with our observations 
and the 20 cm fluxes reported in the public catalogues (NVSS and 
FIRST), their values should be treated as approximate since the 
images used to measure the fluxes were not matched in resolution. 
We also caution that variability among the radio AGN is a legitimate 
concern and could lead to inaccurate spectral index estimates 
since the two radio bands were observed several years apart.  

\begin{figure}
\centerline{
\epsfig{file=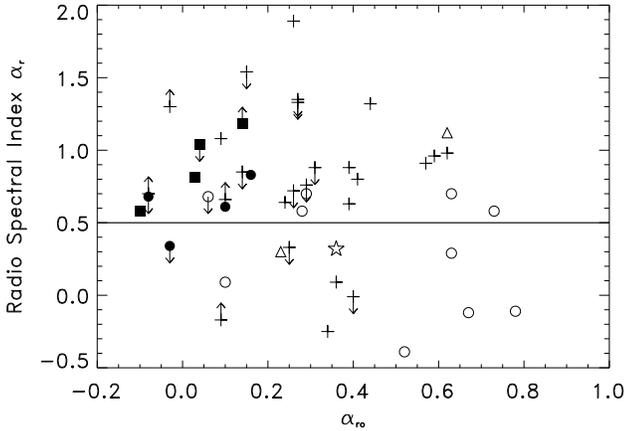, width=9cm}
}
\caption{ The radio spectral index as function of the two point 
spectral index $\alpha_{ro}$ for the 50 HELLAS sources for which 
an information on the radio spectral index is available.   
 Symbols as in Figure~\ref{Rmag_f6cm}. 
}
\label{Alphar_Aro}
\end{figure}

All the radio spectral index $\alpha_r$ are reported in column X 
of Table 1, while in Figure~\ref{Alphar_Aro} we plot  $\alpha_r$ as function 
of the two point spectral index $\alpha_{ro}$. The horizontal dashed line 
denotes the typical spectral index, $\alpha_r$=0.5, separating 
steep and flat spectrum radio sources. Steep spectral indices 
($\alpha_r>$0.5) often indicate radio emission from star formation or 
lobe-dominated AGN, 
while flat ones ($\alpha_r<$0.5) often indicate emission from 
core-dominated AGN (e.g. Kellermann and Owen 1988).  

A statistical analysis using both the measured values and the 
limits to $\alpha_r$ has been performed with the software package ASURV,  
which implements the methods described by Feigelson \& Nelson (1985) and 
Isobe, Feigelson \& Nelson (1986). We divided our sources 
according to their optical 
classification studying the mean radio spectral index of sources with 
broad emission lines (AGN1, open circles in  Figure~\ref{Alphar_Aro}) 
and with narrow emission lines (AGN2 plus
Emission Line Galaxies, filled symbols in Figure~\ref{Alphar_Aro}). 
We found that while the Type 1 AGNs have a mean radio spectral index 
of $\alpha_r=0.25\pm$0.10, the Type 2 AGNs plus the Emission Line Galaxies 
have a steeper radio spectral index  $\alpha_r=0.69\pm$0.11. 
This result is in agreement with the idea that the radio emission from 
Type 1 AGNs is core dominated  with a flat radio 
spectral index due to 
the self-absorption process. On the other hand, in Type 2 AGN 
and Emission Line 
Galaxies the  nonthermal  radio emission (either due to a 
compact nucleus or to the integrated 
emission arising from the supernovae remnants in a starburst region)
takes place on larger physical scale without self-absorption. 
For the case of the star formation these scales correspond approximately to 
0.1-1.0 Kpc as observed in the local ELGs population (Condon 1989, 
Condon 1992) while for the core dominated sources these scales correspond 
to few parsec as well-know from  many  different radio surveys (see 
Nagar et al. 2000,2002 and Giovannini et al. 2001 for the most recent works). 
Under this hypothesis, we expect a more extended radio emission 
from the HELLAS Type 2 AGNs and ELGs in comparison to the radio 
emission from the HELLAS Type 1 AGNs. Sub-arcsecond radio observations 
are needed to test this hypothesis. 

\subsection{Radio Quiet and Radio Loud AGNs in the HELLAS 
spectroscopically identified subsample} 

Starting from the 61 spectroscopically identified sample of 118 HELLAS sources 
published by LF02 we  selected a subsample of 46 AGNs (37 Type 1 + 
9 Type 2) brighter than 5 $\times$ 10$^{-14}$ \cgs in the 5-10 keV band 
and brighter than R=21 and R=19 (for Type 1 and Type 2 AGNs) in the 
optical band. Radio data are available for all the 46 HELLAS AGNs. 
Using the two point spectral index $\alpha_{ro}$=0.35 we divided the 
sources in RL and RQ. We have 6 ($\sim$ 13\%) RL sources, all 
classified as Type 1 AGN. 

This fraction of RL AGNs is well in agreement with the predictions of 
Della Ceca et al. (1994). Using the de-evolved X-ray Luminosity function 
for  RL and  RQ AGNs they predicted a fraction of RL AGNs of $\sim$10\% 
for an X-ray flux limit of $\sim 5 \times 10^{-14}$ \cgs (see their Figure 11).

\section{Summary and Conclusion} 

We present the results of a complete radio follow-up down to 
a 6 cm flux limit of about 0.3 mJy (3 $\sigma$) obtained 
with the VLA and ATCA radio telescopes of all the 
147 X-ray sources detected in the $Beppo$SAX HELLAS survey. 
Our major conclusions are as follows. 

$\bullet$  From a statistical analysis based on the X-ray, optical and 
radio position, we found 53 X-ray/radio associations over 
a total of 147 sources. We estimated that about 10\% 
($i.e.$ $\sim$ 5 sources) of the proposed associations 
may be spurious and therefore the percentage of the ``real'' 
X-ray/radio associations is in the range $\sim$33-36\%. 
This percentage is in agreement to that obtained in other 
X-ray surveys selected in the hard X-ray band and is significantly 
greater than the percentage found in surveys selected 
in the soft X-ray band. This is probably due to 
a combination of two effects: the availability of deeper radio 
data in comparison to the X-ray flux limit and a selection in harder 
X-ray band. Obscured AGNs are in fact missed in soft X-ray selected samples,  
while they are detected in the radio and hard X-ray 
bands since both these wavelengths are transparent against 
obscuring material. This change in the percentage and nature of the 
radio/X-ray associations that we start to see in the HELLAS survey is 
much more evident in the new deep X-ray $Chandra$ samples. 
Radio observations at the micro Jy level of the  $Chandra$ samples showed, 
in fact, that the radio/X-ray associations in these samples are dominated 
by normal galaxies and starburst at z$\sim$ 0.1-1.3. 

$\bullet$ The value of $\alpha_{ro}=0.35$ used 
to separate the radio-loud and radio-quiet sources  
has been estimated using the Lee statistic.
Of the 53 HELLAS radio/X-ray associations, 26 have been classified 
as radio-loud sources.
From the analysis of the ratio between the radio and the X-ray fluxes, we have 
assumed  that all the 63 sources for which it was impossible to 
calculate the two point spectral index $\alpha_{ro}$ (and than 
to obtain a classification in RQ or RL sources) are likely 
to be radio-quiet objects.  
Under this hypothesis, the fraction of radio-loud objects 
in the HELLAS sample is $\sim$18 per cent. 

$\bullet$ The analysis of the multiband photometric data 
based on the two point spectral indices $\alpha_{ro}$ and 
$\alpha_{ox}$ has shown that the 
identified radio-loud objects are associated mainly with Type 1 AGN
with L$_{5-10~keV} \gax 10^{44}$ erg/s,   
while the identified Type 2 AGNs and Emission Line Galaxies are 
radio-quiet sources with L$_{5-10~keV} \lax 10^{44}$ erg/s. 
RQ Type 1 AGNs and Type 2 AGNs cannot be 
distinguished only on the basis of their photometric data. On the basis 
of these results we would conclude that all the unidentified sources 
with  $\alpha_{ro}$ $\gax$ 0.30 - 0.35 and $\alpha_{ox} \lax$ 1.5 
are likely to be Type 1 AGN. 

$\bullet$ The analysis of radio spectral index has shown that Type 1 
AGNs have a mean  spectral index flatter than Type 2 AGNs and Emission 
Line Galaxies. 
This result is in agreement with the idea that 
the core-dominated radio emission from 
Type 1 AGNs is self-absorbed, while 
in AGN2 and Emission Line Galaxies the radio emission take place on 
larger physical scale, without self-absorption. 

$\bullet$ Finally, using a subsample of 46 AGNs (37 Type 1 and 9 Type 2, 
see section 4.4) we studied the 
fraction of the RL and RQ AGNs. We have 6 ($\sim$ 13\%) RL sources , all 
classified as Type 1 AGN. This fraction of RL AGNs is well 
in agreement with the results of 
Della Ceca et al. 1994 which predicted a fraction of RL AGNs of $\sim$10\% 
for an X-ray flux limit of $\sim 5 \times 10^{-14}$ \cgs.

\vspace{1cm}

\centerline{\em Acknowledgements}

This paper is based on observations collected at the Very Large Array (VLA) 
Radio Telescope and at the  Australia Telescope Compact Array (ATCA). 
The VLA is a facility of National Radio Astronomy Observatory (NRAO) which 
is operated by Associated Universities, Inc., under cooperative 
agreement with the National Science Foundation. 
The ATCA is part of the Australia Telescope which is funded by the 
Commonwealth of Australia for operation as a National Facility 
managed by CSIRO.
This work was supported by the Italian Ministry for University 
and Research (MURST) under grant COFIN-00-02-36 
and by Italian Space Agency ASI contracts I/R/113/01 and 
I/R/073/01.  CV also acknowledges the NASA LTSA grant NAG5-8107 
for financial support.

\end{document}